\documentstyle[aps,prl]{revtex}
\twocolumn
\begin{document}
\tolerance=10000
\draft
\title{Surface critical behavior in fixed dimensions $d<4$: Nonanalyticity of
critical surface enhancement and massive field theory approach}
\author{H.~W.~Diehl and M.\ Shpot\cite{permadmyk}}
\address{
Fachbereich Physik, Universit\"at - Gesamthochschule - Essen,\\
D-45117 Essen, Federal Republic of Germany
}
\date{August 15, 1994}
\maketitle
\begin{abstract}
The critical behavior of semi-infinite systems in fixed dimensions $d<4$
is investigated theoretically.
The appropriate extension of Parisi's massive field theory approach is
presented.
Two-loop calculations and subsequent Pad\'e-Borel analyses
of surface critical exponents of the special and ordinary phase transitions
yield estimates in reasonable agreement with recent Monte Carlo results.
This includes the crossover exponent $\Phi (d=3)$, for which we obtain
the values $\Phi (n=1)\simeq 0.54$ and $\Phi (n=0)\simeq 0.52$,
considerably lower than the previous $\epsilon$-expansion estimates.
\end{abstract}

\narrowtext
Methods of field theory have become an essential ingredient of the modern
theory of critical phenomena, providing both a conceptually appealing
theoretical framework as well as powerful tools for quantitative analyses.
Many important field-theoretical techniques, developed and tested originally
in the study of {\it bulk} critical behavior,
subsequently have been extended
to systems with boundaries and utilized with
considerable success in the analysis
of {\it surface} critical phenomena at bulk critical points \cite{hwd}.

One important method that, to our knowledge, has not yet been employed in the
study of surface critical phenomena is the {\it field-theoretic renormalization
group (RG) approach in a fixed space dimension} $d$ below the upper critical
dimension $d^*$ \cite{Parisi}. The merits of this approach are well known:
pushed to sufficiently high orders of perturbation theory
and combined with sophisticated series
resummation techniques, it has produced \cite{bnm78,lgz80} values of bulk
critical exponents belonging to the most accurate
ones obtained so far \cite{LGZJ,Pawley,liufish}. In addition, it has
proved very useful in quantitative investigations of preasymptotic critical
behavior \cite{bb85,bbmn,SD,hald}.

In the present Letter we show how this approach can be generalized
to systems with surfaces. Aside from the fundamental importance of such an
extension, we have been motivated by the discrepancies existing between
the results of recent Monte Carlo analyses
for the crossover surface exponent
$\Phi$ in three-dimensional systems \cite{ml88,lb90,rdww,hg94}
and previous numerical estimates of $\Phi$
obtained by extrapolation of its second-order $\epsilon=4-d$ expansion to
$\epsilon=1$ \cite{hwd}. In particular, the most recent Monte Carlo
values for $\Phi$ of both the Ising \cite{rdww}
and polymer \cite{hg94} universality classes are
{\it substantially lower} than the original $\epsilon$-expansion
estimates, and also {\it significantly lower} than previous results of
computer simulations \cite{ekb,ml88,lb90}.
The annoyingly large discrepancy with the $\epsilon$-expansion results
calls for clarification and improvement of field-theoretic analyses.
With this in mind, we have extended the massive field theory approach
\cite{Parisi} to the study of surface critical behavior.
Using this approach, we have performed two-loop calculations
for the semi-infinite $\phi^4_{d=3}$ model. Subsequent
Pad\'e-Borel analyses of the resulting series expansions
yield the estimates for surface critical exponents shown
in Table 1.

The Hamiltonian of the model is
\begin{eqnarray}\label{Ham}
{\cal H}&=&\int_V\!d^dx\left( \case{1}/{2}(\nabla\bbox{\phi})^2+
\case{1}/{2}\tau_0\,\bbox{\phi} ^2+
\case{1}/{4!}u_0\,|\bbox{\phi}|^4\right)
\nonumber\\&&\mbox{}
+\int_{\partial V}\!d^{d-1}x_{\parallel}\,
\case{1}/{2}\,c_0\,\bbox{\phi}^2\;,
\end{eqnarray}
where $\bbox{\phi}=(\phi_\alpha({\bf x}))$ is an $n$-component field
on the half-space $V=\{{\bf x}=({\bf x}_\parallel ,z)
\mid {\bf x}_\parallel\in
I\!\!R^{d-1}, z\ge 0\}$ bounded by $\partial V$, the $z=0$ plane. For the
time being we assume the regularization of parallel-momentum integrations
%that the ultraviolet (uv) singularities have been regularized by means of
by a large-momentum cutoff $\Lambda$.

In wishing to study critical systems in fixed dimensions $d<d^*$
by means of perturbative methods, one is faced with the
familiar occurrence of infrared (ir) singularities.
Feynman integrals involving
massless propagators become ill-defined.
This is a problem typical of massless {\it super}renormalizable
field theories. Nevertheless such theories exist,
owing to the appearance of a nonperturbative critical bare mass shift
\cite{Parisi,Sym,bergere}. In the dimensionally regularized theory, this
shift corresponds to a critical value of $\tau_0$ of the form
$\tau_{0c}=u_0^{2/\epsilon}\>{\cal T}(\epsilon)\;$, where
${\cal T(\epsilon)}$ has simple poles at $\epsilon= \epsilon_k\equiv
2/k,\,k\in I\!\!N$.

In the case of our semi-infinite model (\ref{Ham}),
an independent mass scale is provided by $c_0$, the bare
surface enhancement. Its physical significance is
well-known \cite{hwd}: depending on whether $c_0$ is smaller, equal,
or larger than a certain critical value
$c_{\text{sp}}=c_{\text{sp}}(\epsilon,u_0,\Lambda)$,
distinct types of surface transitions
occur at the bulk critical point, called ordinary, special, and extraordinary
transitions, respectively. Finite values of $c_{\text{sp}}$
exist, of course, only if $d$ exceeds the lower critical dimension
$d_*(n)=3-\delta_{n1}$ for the occurrence of surface ordering.

The quantity $c_{\text{sp}}$ has a number of properties
analogous to $\tau_{0c}$. First, in the cut-off regularized theory,
it diverges as $\Lambda\to\infty$. Its leading uv singularity for $d<4$
is of the form $u_0\,\Lambda^{d-3}$, similar as  $\tau_{0c}$ has
one $\sim u_0\,\Lambda^{d-2}$.

A second analogy becomes apparent if
we consider the uv behavior of correlation functions.
In the case of the $\phi^4_{d<4}$ bulk theory (with $V=I\!\!R^d$),
a mass shift $\tau_0=\tau_{0c}+\delta\tau_0$ is known to be sufficient
to absorb the uv singularities of the $N$-point vertex and correlation
functions. In our semi-infinite case we must keep track in correlation
functions of which points are taken {\it on} or {\it off} the surface. Let
$G^{(N,M)}(\{{\bf x}_i\},\{{\bf r}_j\}; \tau_0, u_0, c_0,\Lambda)$ be the
(regularized) connected $(N+M)$-point correlation functions
involving $N$ fields $\phi({\bf x}_i)$ at distinct points
${\bf x}_i,\,1\le
i\le N$, off the surface and $M$ fields
$\phi ({\bf r}_j,z=0)\equiv \phi_s({\bf r}_j)$ at distinct
surface points with parallel coordinates ${\bf r}_j,\, 1\le j\le M$.
To absorb the uv singularities of these functions for $d<4$ both a
mass shift and a surface-enhancement shift $c_0=c_{\text{sp}}+\delta c_0$
is required; i.e., the limits $\Lambda\to\infty$ of the $G^{(N,M)}$ with
fixed $\delta\tau_0>0$ and $\delta c_0>0$ exist.

A further analogy is the form of $c_{\text{sp}}$ in the dimensionally
regularized theory. On dimensional grounds one expects it to read
\begin{equation}\label{csp}
c_{\text{sp}}=u_0^{1/\epsilon}\>{\cal C}(\epsilon)\;,
\end{equation}
where ${\cal C}(\epsilon)$ has poles at the above values $\epsilon_k$.
The value of $c_{\text{sp}}$ is fixed by
the requirement that the layer susceptibility
$\chi_{11}=\hat{G}^{(0,2)}({\bf p}=0)$ diverges at the special point
$(\tau_{0c},c_{\text{sp}})$.
Here $\hat{G}^{(0,2)}({\bf p})$ is the parallel Fourier transform
of %the surface pair correlation function
$G^{(0,2)}({\bf r}-{\bf r}')$.
Hence $c_{\text{sp}}$ is the solution to
$\chi_{11}^{-1}(\tau_{0c},u_0,c_0=c_{\text{sp}})=0\;$.
By explicit computation of
the $O(u_0)$ term in the expansion of $\chi_{11}^{-1}$ one
may verify that it has indeed a pole at $d=3$,
which must be eliminated by a surface-enhancement
shift.	Thus, just as in the bulk theory, one encounters the following
situation: although, by making appropriate shifts of $\tau_0$
and $c_0$, the theory can be
made uv-finite for $d<4$, the critical values $\tau_{0c}$ and
$c_{\text{sp}}$ {\it cannot} be determined
by perturbation theory owing to their
nonanalytic dependence on $u_0$. In order to obtain meaningful results
in spatial dimensions $d$ with $d_*<d<d^*$ one should employ an
appropriate extended version of the massive field theory RG approach,
avoiding direct computations of both $\tau_{0c}$ and  $c_{\text{sp}}$.

To explain our method, we start by recalling from
Ref.\ \onlinecite{hwd} that the field $\phi({\bf x})$
and the surface field $\phi_s({\bf r})$ should be reparametrized by distinct
renormalization factors. Accordingly we introduce the renormalized
correlation functions through
\begin{equation} \label{gnm}
G_{\text{ren}}^{(N,M)}(.;m,u,c)=Z_\phi^{-(N+M)/2}\,
Z_1^{-M/2}\,G^{(N,M)}\;.
\end{equation}
Unlike the renormalized surface enhancement $c$ and
the renormalization factor $Z_1$ (which are specific to our
surface-bounded system),
the renormalized mass $m$, renormalized coupling constant $u$, and
$Z_\phi$ are bulk quantities. We fix
the latter ones by standard normalization conditions for the renormalized
bulk vertex functions $\Gamma_{b,\text{ren}}^{(N)}$. In terms of their
Fourier transforms $\tilde{\Gamma}_{b,\text{ren}}^{(N)}(\{{\bf q}\})$ these
conditions read \cite{Parisi}
\begin{equation}\label{normm}
\tilde\Gamma_{b,\text{ren}}^{(2)}({\bf 0};m,u)=m^2,\quad
\end{equation}
\begin{equation}\label{normZphi}
{\partial \over\partial q^2}\left.\tilde{\Gamma}_{b,\text{ren}}
^{(2)}({\bf q};m,u)
\right|_{q^2=0}=1,
\end{equation}
\begin{equation}\label{normu}
\tilde{\Gamma}^{(4)}_{b,\text{ren}}(\{{\bf 0}\};m,u)=m^\epsilon\, u\;.
\end{equation}

In order to specify $c$ and $Z_1$,
we require that
\begin{equation}\label{normc}
\hat{G}_{\text{ren}}^{(0,2)}({\bf p};m,u,c)\big|_{{\bf p}={\bf 0}}=
{1\over m+c}
\end{equation}
and
\begin{equation}\label{normZ1}
{\partial\over\partial p^2}\,\left.\hat{G}_{\text{ren}}^{(0,2)}({\bf p};m,u,c)
\right|_{p^2=0}=-{1\over 2m(m+c)^2}\;.
\end{equation}
These normalization conditions are suggested by the familiar
zero-loop expression \cite{hwd} $[c_0+(p^2+\tau_0)^{1/2}]^{-1}$ for
$\hat{G}^{(0,2)}({\bf p})$. Together with (\ref{normm}), the first one,
(\ref{normc}), ensures
that the special point
is located at $m=c=0$.  Equation (\ref{normZ1}) fixes $Z_1$ in much
the same way as (\ref{normZphi}) determines $Z_\phi$.

Since we are also interested in the crossover exponent $\Phi$, we should
also consider correlation functions with insertions of the surface operator
$\phi_s^2$. Let $G^{(N,M;I,I_1)}$ denote the connected correlation function
with $N$ fields $\phi$ off the surface, $M$ surface fields $\phi_s$, $I$
insertions of $\case{1}/{2}\phi^2$ at points off the surface,
and $I_1$ insertions of $\case{1}/{2}\phi_s^2$. We write the renormalized
operators as $(\phi^2)_{\text{ren}}=Z_{\phi^2}\,\phi^2$ and
$(\phi_s^2)_{\text{ren}}=Z_c\,\phi_s^2$. The bulk factor $Z_{\phi^2}$ we fix
through the usual normalization condition
$\tilde{\Gamma}_{b,R}^{(2,1)}(\{{\bf 0}\})=1$
for the Fourier transform of the bulk two-point vertex function with an
insertion of $\case{1}/{2}\phi^2$. To specify $Z_c$ we require that
\begin{equation}\label{normZc}
 \hat{G}_{\text{ren}}^{(0,2;0,1)}(\{{\bf p}\};m,u,c)
\big|_{\{{\bf p}={\bf 0}\}}= (m+c)^{-2}\;.
\end{equation}
Again, this condition is suggested by the zero-loop result for
$\hat{G}^{(0,2)}$, as can be seen by comparison with (\ref{normc}), noting
that the bare analog of the left-hand side of (\ref{normZc}) may be written as
$(-\partial/\partial c_0)\,\hat{G}^{(0,2)}({\bf 0})$.

The above normalization conditions determine
$\tau_0$, $u_0$, $Z_\phi$, $Z_{\phi^2}$, $c_0$, $Z_1$, and $Z_c$
as functions of $u$, $c$, and $\Lambda$. All
$Z$-factors have a finite $\Lambda\to\infty$
limit in the $d<4$ case considered here. Let us set $\Lambda=\infty$ in the
sequel. (In our calculations we actually took $\Lambda=\infty$ from the
outset, employing dimensional regularization.)
Then the bulk $Z$-factors $Z_\phi$ and $Z_{\phi^2}$
become functions of the single dimensionless variable $u$.
However, with our choice of normalization conditions,
the surface $Z$-factors $Z_1$ and $Z_c$
depend on both $u$ and the dimensionless ratio $c/m$.

By varying $m$ at fixed $u_0$, $c_0$, and $\Lambda$,
the analogs of the Callan-Symanzik equations (CSE) can be
derived in a straightforward fashion. Let us define
the functions $\beta(u)=m\partial_m|_0\,u$,
$B_c(u,c/m)\equiv m\partial_m|_0\ln c$, $\eta_\phi(u)\equiv
m\partial_m|_0\,\ln Z_\phi$,
and $\eta_1(u,c/m)\equiv m\partial_m|_0\,\ln Z_1$, where $|_0$
indicates that the derivatives are taken at fixed $u_0$, $c_0$,
and $\Lambda$. Upon introducing the differential operator
\begin{equation}\label{Dm}
{\cal D}_m=m\,\partial_m+\beta\,\partial_u+\,B_c\,c\,\partial_c
+{{N+M}\over 2}\,\eta_\phi+{M\over 2}\,\eta_1\;,
\end{equation}
the CSE of the $G_{\text{ren}}^{(N,M)}$ can be written as
\begin{equation}\label{CSE}
{\cal D}_m\,G_{\text{ren}}^{(N,M)}=-(2-\eta_\phi)\,m^2\,\int_V
d^dX\,G_{\text{ren}}^{(N,M;1,0)}\;,
\end{equation}
where the integration is over the position ${\bf X}$ of the inserted
$\phi^2$ operator.

Just as in the bulk case, and as could be corroborated by means of a
short-distance expansion, the inhomogeneity on the right-hand
side should be negligible in the critical regime.
The resulting homogeneous CSE differs from its standard bulk analog in
that it involves functions $\beta_c$ and $\eta_1$
of two variables, $u$ and $c/m$. In a study of the
crossover from the critical behavior
characteristic of the special transition (for $c/m\ll 1$)
to that of the ordinary transition (for $c/m\gg 1$) it would be
essential to carry along this dependence on $c/m$.
Such a crossover analysis is beyond the scope of this Letter.
Our main aim here is the calculation of the surface
critical exponents of these two surface transitions.
Therefore we focus directly on the respective asymptotic behavior.

The easier case is the special transition, which we consider first.
Its critical behavior can be investigated upon setting $c=0$.
The homogeneous CSE can then be integrated in a standard fashion.
One finds that the surface correlation exponent can be written as
\begin{equation} \label{etapar}
\eta_\parallel^{\text{sp}}=\eta_1^{\text{sp}}(u^*)+\eta\;,
\end{equation}
where $\eta \equiv\eta_\phi(u^*)$ is the standard bulk correlation
exponent, $u^*$ is the ir-stable zero of $\beta(u)$, and
$\eta_1^{\text{sp}}(u)\equiv\eta_1(u,0)$.
A similar analysis applied to the CSE for correlation functions
with insertions of $\phi^2_s$
shows that the anomalous dimension of this operator is given by the
fixed-point value of the function
$\eta_c(u,c/m)\equiv m\partial_m|_0\,\ln Z_c$.
Hence the surface crossover exponent can be expressed as
\begin{equation} \label{Phi}
\Phi=\nu [1+\eta_c^{\text{sp}}(u^*)]
\end{equation}
in terms of $\eta_c^{\text{sp}}(u)\equiv\eta_c(u,0)$ and
the bulk exponent $\nu$.

We have computed the perturbation expansions of the functions
$\eta_\parallel^{\text{sp}}(u)\equiv\eta_1^{\text{sp}}(u)+\eta_\phi(u)$
and $\eta_c^{\text{sp}}(u)$ to
two-loop order for $d=3$. Let us make the usual change of normalization of
$u$ such that
$\beta(u)=-u+u^2+O(u^3)$. Then our results become
%\widetext
\begin{eqnarray}\label{etapsp}
\eta_\parallel^{\text{sp}}(u)
&=&-{n+2\over 2(n+8)}u+
\left.{12\,(n+2)\over (n+8)^2} \right[ 2A
\nonumber\\ &&\mbox{}
+\left.{n+2\over6}\left( {1\over 2}-\ln2+\ln^2 2\right)
-{n+6\over16}\right]u^2
\end{eqnarray} and
\begin{eqnarray}
\eta_c^{\text{sp}}(u)&=&-{n+2\over n+8}\left(\ln 4-{1\over2}\right)u
\nonumber\\ &&\mbox{}
- {24\,(n+2)\over (n+8)^2} \left[ A-B
+{n+8\over12}\left(\ln2-{1\over4}\right)\right.
\nonumber\\ &&\mbox{}
+{n+2\over3}
\left.\left(\ln^2 2-{7\over4}\ln2+{19\over32}\right)\right] u^2\;.\label{etac}
\end{eqnarray}
%\narrowtext\noindent
The numbers $A$ and $B$ are associated with contributions of
two-loop diagram with three equivalent internal lines. We have determined them
by numerical integration, obtaining
$A=0.202428$ and $B=0.678061$.

{}From these results analogous series expansions
can be derived for other surface exponents
by means of scaling laws \cite{hwd}. The coefficients of most
of the resulting series alternate in sign and decrease in
absolute value \cite{tbp}.
Exceptions are some series involving $\eta_c^{\text{sp}}$,
whose behavior is rather bad. We have performed Pad\'e analyses and
Pad\'e-Borel resummations of the direct and
inverse series for the indices
$\eta_\parallel,\Delta_1,\eta_\perp,\beta_1,
\gamma_{11},\gamma_1,\delta_1,\delta_{11}$, $\alpha_1,
\alpha_{11}$, and $\Phi$. In these computations we have used the fixed-point
values $u^*(n=0)$ = 1.632 and $u^*(n=1)$ = 1.597 from the
Pad\'e-Borel calculations of Refs.\,\cite{bgmn} and \cite{jug}.

In the group of exponents $\eta_\parallel,\ldots,\delta_{11}$
related to $\eta_1^{\text{sp}}(u^*)$ the most reliable estimate is obtained
for $\Delta_1$, which appears to exhibit the best convergence
properties. Pad\'e-Borel resummations of this series yield the values
$\Delta_1(n=0)=0.921$ and $\Delta_1(n=1)=0.997$.
The estimates in Table\ 1 for the other exponents of
this group have been derived from these values of $\Delta_1$ via scaling
relations, using the $d=3$ values of $\nu$ and $\eta$
given in Ref.\ \onlinecite{lgz80}.

In the group of exponents $\alpha_1,\,\alpha_{11},\,\Phi$
related to $\eta_c^{\text{sp}}(u^*)$ the series
for $\alpha_1$ yields estimates
with the least scattering. The values of $\alpha_{11}$ and $\Phi$
listed in Table 1 have been obtained from our
results for $\alpha_1$ and the accepted
bulk values of $\nu$ and $\eta$.

In order to study the ordinary transition, we must consider the limit
$m\to 0$ at fixed $c>0$; i.e., $c/m\to\infty$. A well-known complication
is that one cannot simply set $c=\infty$ in the functions $G^{(N,M)}$ with
$M>0$ because a Dirichlet boundary condition holds for $c=\infty$ \cite{hwd}.
Yet, the cumbersome calculation of $c$-dependent quantities can be avoided
by considering correlation functions involving instead of $\phi_s$ the normal
derivative $\partial_n\phi$ at the surface. Let $G_\infty^{(N,M)}$ be the
analog of $G^{(N,M)}$  resulting by replacement of all the $\phi_s$ by
$\partial_n\phi$, with $c_0=\infty$. Following the strategy described in
Ref.\ \onlinecite{hwd}, we have performed an independent RG analysis of
these functions. The renormalized operator
$(\partial_n\phi)_R=(Z_\phi Z_{1,\infty})^{-1/2}\,\partial_n\phi$
involves a renormalization factor $Z_{1,\infty}(u)$, which we fix
through the condition
$(\partial/\partial p^2) \hat{G}_{\infty,R}^{(0,2)}({\bf p},m,u)
 \big|_{p^2=0}=-1/2m$ \cite{tbp,remark}.

The associated RG function
$\eta_{1,\infty}(u)\equiv m\, \partial_m|_0 \ln Z_{1,\infty}$
gives us the function
\begin{equation}
\eta_{\parallel}^{\text{ord}}(u)=2+\eta_{1,\infty}(u)+\eta_\phi(u)
\end{equation}
needed to determine the surface exponent
$\eta_\parallel^{\text{ord}}=\eta_{\parallel}^{\text{ord}}(u^*)$.
The explicit two-loop expression of this function for $d=3$
reads
\begin{equation}
\eta_\parallel^{\text{ord}}(u) =2-{n+2\over 2(n+8)}\, u
- {24\,(n+2)\over (n+8)^2}\left[C+ {n+14\over 96}\right] u^2
\label{eta1infty}
\end{equation}
with $C=-0.105063$.

Again, this result can be combined with scaling laws to obtain
the corresponding expansions of the other surface critical indices of the
ordinary transition. In most cases the coefficients of these series do not
alternate in sign, with similar behavior of the inverse ones. Therefore,
they are not adapted to Pad\'e-Borel resummation. The numerical
values of the ordinary-transition exponents arising from $[1/1]$
Pad\'e approximants for each individual series are listed in Table 1. The
exception is $\gamma^{\text{ord}}_{11}$ for which
this approximant does not exist. We evaluated this index using the
values of $\eta_\parallel^{\text{ord}}$ and $\nu$. For $n=2$ and $n=3$,
the fixed-point values $u^*=1.558$ and $u^*=1.521$ \cite{jug} were used,
respectively.

Our numerical values of the surface critical exponents gathered in Table 1
generally are in reasonable agreement both with previous estimates
based on the $\epsilon$
expansion \cite{hwd} as well as with those obtained by
other means (see Refs.\cite{hwd} and
\cite{ml88,lb90,rdww,hg94,ekb,rugedis} for comparisons).
Note, however, that our estimates
for $\Phi$ are definitively lower than the values
$\Phi(n=1)\simeq 0.68$ and $\Phi(n=0)\simeq 0.67$
quoted in Ref.\ \onlinecite{hwd},
which were
obtained by setting $\epsilon=1$ in the $\epsilon$ expansion of $\Phi$
to order $\epsilon^2$.
Recent Monte Carlo simulations have yielded the significantly lower
estimates $\Phi(1)=0.461\pm0.015$ \cite{rdww}, $\Phi(0)=0.530\pm0.007$
\cite{ml88}, and $\Phi(0)=0.496\pm0.005$ \cite{hg94}. Our results
$\Phi(1)\simeq 0.54$ and $\Phi(0)\simeq 0.52$ are fairly close to
these values. This indicates that the
crossover exponents $\Phi$ for $d=3$ and $n=0,1$ are indeed smaller
than previously thought. Let us also note that when the $\epsilon$
expansion to order $\epsilon^2$ is extrapolated to $d=3$ by means of more
elaborate techniques (e.g, Pad\'e-Borel resummation), estimates
for $\Phi(0)$ and $\Phi(1)$ as low as ours can be obtained.

A more detailed account of our work will be given elsewhere \cite{tbp}.

This work has been made possible through a research fellowship for M.\ S.\
by the Humboldt foundation and through partial support by the Deutsche
Forschungsgemeinschaft through SFB 237. It is our pleasure to thank both
organisations.

\begin{table}[hbtp]
\caption{Estimated $d=3$ values of the surface critical exponents of the
special and ordinary transition for various values of $n$.}
\begin{tabular}{ldddddd}
&\multicolumn{2}{d}{SPECIAL}& \multicolumn{4}{d}{ORDINARY}\\
\hline
 & $n=0$ & $n=1$ & $n=0$ & $n=1$ & $n=2$ &$n=3$\\
\hline
 $\eta_{\parallel}$&$-$0.133& $-$0.165& 1.660&1.528 &1.422 &1.338\\
$\Delta_1$&  0.921& 0.997&0.382 & 0.450&0.514 &0.574\\
$\eta_{\perp}$&$-$0.053& $-$0.067 &0.859 &0.801 &0.753 &0.714\\
$\beta_1$&0.255&0.263 & 0.817& 0.845& 0.868&0.889\\
$\gamma_{11}$&0.666&  0.734& $-$0.388& $-$0.333& $-$0.282&$-$0.238\\
$\gamma_1$&1.207& 1.302& 0.664& 0.742& 0.815&0.882\\
$\delta_1$&5.734& 5.951& 1.854& 1.941& 2.019&2.088\\
$\delta_{11}$&3.612& 3.791& 0.504& 0.582& 0.651&0.711\\
$\alpha_1$&0.342  &0.279 & & & &\\
$\alpha_{11}$&$-$0.140 &$-$0.182& & & &\\
$\Phi $&0.518  &0.539 & & & &
\end{tabular}
\end{table}
\end{document}